\documentclass[%
 reprint,
 amsmath,
 amssymb,
 nofootinbib,
 aps,
 prd
]{revtex4-2}
\usepackage{amsmath}
\usepackage{amssymb}
\usepackage{graphicx}
\usepackage{dcolumn}
\usepackage{bm}
\usepackage[nolist, nohyperlinks]{acronym}
\usepackage{hyperref}

\usepackage{xcolor}
\usepackage{soul}

\begin{document}

\preprint{}

\title{Spectra for Reactions in Astrophysical Electromagnetic Cascades with Lorentz Invariance Violation: The Vacuum Cherenkov Effect}

\author{Andrey Saveliev}
\email{anvsavelev@kantiana.ru}
\affiliation{Immanuel Kant Baltic Federal University, Ul. A. Nevskogo 14, 236016 Kaliningrad, Russia}
\affiliation{Lomonosov Moscow State University, GSP-1, Leninskiye Gory 1-52, 119234 Moscow, Russia}

\author{Rafael \surname{Alves Batista}}
\email{rafael.alves\_batista@iap.fr}
\affiliation{Sorbonne Université, Institut d’Astrophysique de Paris (IAP), CNRS UMR 7095 98 bis bd Arago 75014, Paris, France}
\affiliation{Sorbonne Université, Laboratoire de Physique Nucléaire et de Hautes Energies (LPNHE), 4 place Jussieu, F-75252, Paris, France}

\author{Feodor Mishin}
\email{20mishinf@harrowschool.org.uk}
\affiliation{Harrow School, 5 High St, Harrow HA1 3HP, United Kingdom}

\begin{abstract}
Lorentz invariance violation is a feature of several quantum gravity models in which Lorentz symmetry is broken at high energies, leading to potential changes in particle behavior and interactions. In this study, we investigate vacuum Cherenkov radiation, a reaction in which an electron spontaneously emits a photon. This process, forbidden when considering unbroken Lorentz symmetry, is a phenomenological consequence of some quantum gravity models. We derive, for the first time, the spectra for the vacuum Cherenkov reaction, and confirm our results numerically. These results can be used to derive limits on Lorentz invariance violation.
\end{abstract}

\maketitle

\acrodef{AGN}{active galactic nucleus}
\acrodefplural{AGN}{active galactic nuclei}
\acrodef{A.U.}{astronomical unit}
\acrodef{BSM}{beyond the Standard Model}
\acrodef{CMB}{cosmic microwave background}
\acrodef{CPT}{charge, parity, and time} 
\acrodef{CRB}{cosmic radio background}
\acrodef{CTA}{Cherenkov Telescope Array}
\acrodef{DSR}{deformed (or doubly) special relativity}
\acrodef{EBL}{extragalactic background light}
\acrodef{GR}{general relativity}
\acrodef{GRB}{gamma-ray burst}
\acrodef{HE}{high-energy}
\acrodef{IACT}{imaging air-Cherenkov telescope}
\acrodef{ICS}{inverse Compton scattering}
\acrodef{IGMF}{intergalactic magnetic field}
\acrodef{LHAASO}{Large High Altitude Air Shower Observatory}
\acrodef{LHC}{Large Hadron Collider}
\acrodef{LIV}{Lorentz invariance violation}
\acrodef{mSME}{minimal Standard-Model extension}
\acrodef{PD}{photon decay}
\acrodef{PP}{pair production}
\acrodef{PS}{photon splitting}
\acrodef{QCD}{quantum chromodynamics}
\acrodef{QED}{quantum electrodynamics}
\acrodef{QG}{quantum gravity}
\acrodef{SM}{Standard Model}
\acrodef{SME}{Standard-Model extension}
\acrodef{SR}{special relativity}
\acrodef{SWGO}{Southern Wide-field Gamma-ray Observatory}
\acrodef{UHE}{ultra-high-energy}
\acrodef{UHECR}{ultra-high-energy cosmic ray}
\acrodef{VC}{vacuum Cherenkov}
\acrodef{VHE}{very-high-energy}

\section{Introduction}

The \ac{SM} of particle physics is an exceptionally successful theory that explains the electromagnetic, weak, and strong interactions, covering the majority of observed phenomena in physics. However, several key experimental and theoretical challenges remain unresolved, including the nature of dark matter and dark energy, the origin of neutrino masses, and the hierarchy problem. Additionally, the \ac{SM} does not incorporate gravity, which is a fundamental force in Nature. Addressing this gap has led to numerous formalisms which may be viewed as attempts to create a theory of \ac{QG} which seeks to reconcile the quantum field theory of the \ac{SM} with the differential geometry framework of \ac{GR}. \ac{GR} itself faces issues such as singularities and the black hole information loss problem, further highlighting the need for a unified theory.

Effects of \Acl{QG}~\cite{Addazi:2021xuf} are expected to become more prominent at energies close to or beyond the Planck scale, either at distances close to or smaller than the Planck length ($\lambda_{\rm Pl} = 1.62 \times 10^{-35}\,{\rm m}$) or at energies near or above the Planck mass ($M_{\rm Pl} = 1.22 \times 10^{28}\,{\rm eV}$). These scales are far beyond the reach of current particle accelerators like the \ac{LHC}, which operates at a center of mass energy of 14~TeV, 15 orders of magnitude below the Planck scale. However, specific \ac{QG} tests might involve other physical quantities, such as the energy of a particle in certain reference frames or large cosmological propagation distances. These quantities could offset the Planck scale suppression, making minimal corrections observable.

One possible consequence of such approaches \ac{BSM}, and specifically of QG, is \ac{LIV}~\cite{Collins:2004bp, Alfaro:2004aa, Jacobson2006150, Sotiriou:2009bx, Reyes:2014wna}. This is in particular interesting since \ac{CPT} violation implies \ac{LIV} \cite{Greenberg:2002uu}. A common strategy to implement this \ac{LIV} from the perspective of field theory is to construct a minimal extension of the \acl{SM}, typically by adding extra terms to the \ac{SM} Lagrangian~\cite{Colladay:1998fq,Kostelecky:2018yfa}. This results in an effective field theory that incorporates \ac{LIV}.

In terms of particle dynamics, \ac{LIV} primarily affects particle propagation by altering the dispersion relation, which then takes the form
\begin{equation}
E_{\rm LIV}^{2} = E_{\rm SM}^{2} + f_{\rm LIV}(p)\,,
\label{eq:generalDR}
\end{equation}
where $E_{\rm LIV}$ is the energy of the particle in the presence of \ac{LIV}, $E_{\rm SM}$ is its energy without \ac{LIV} and $f_{\rm LIV}(p)$ is the shift due to \ac{LIV},  usually dominated by a single power of the particle's momentum $p$, i.e.~$f_{\rm LIV}(p) \propto \mathcal{O}(p^{n+2})$ with $n \geq 0$. This modification impacts the reaction thresholds and, consequently, alters the propagation length of particles, as it shifts the limits in the integral related to their calculation. 

\Ac{LIV} allows for new processes that are kinematically forbidden under the usual Lorentz symmetry to take place. The most relevant ones include spontaneous \acl{PD} ($\gamma \rightarrow e^+ + e^-$) and \acl{PS} ($\gamma \rightarrow \gamma + \gamma$), as well as the \acl{VC} ($q^\pm \rightarrow q^\pm + \gamma$) effect for electrons and other charged particles. In addition, multi-nucleon systems such as atomic nuclei can also undergo spontaneous disintegration~\cite{JCAP10-03-046}. For a comprehensive review of how \ac{LIV} modifies processes in electromagnetic cascades and \ac{UHECR} propagation, see \cite{Mattingly:2005re, Galaverni:2008yj, Martinez-Huerta:2020cut} and \cite{Mattingly:2005re, Bietenholz:2008ni, Martinez-Huerta:2017ulw, JCAP10-03-046}, respectively.

These changes could significantly alter the way particles propagate, thus influencing the interpretation of astrophysical measurements. In this work we will consider the impact of \ac{VC} on the propagation of astrophysical electromagnetic cascades which are initiated by a high-energy gamma-ray photon from an astrophysical source. Therefore, our results may be relevant in the future for the interpretation of gamma-ray observations regarding \ac{LIV}, as has been already done for such objects as \acp{AGN}~\cite{hess2011a}, specifically blazars \cite{hess2019a,Levy:2024eiq,MAGIC:2024ifx}, \acp{GRB}~\cite{Amelino-Camelia:1997ieq,hess2011a,magic2020c,Du:2020uev,Finke:2022swf,Piran:2023xfg,LHAASO:2024lub,Liao:2024wbd}, the Crab Nebula~\cite{satunin2019a, astapov2019a, HAWC:2019gui} or PeVatrons \cite{LHAASO:2021opi}. For some recent reviews of experimental searches for \ac{LIV},  in particular in the photon sector, see, for example, Refs.~\cite{Lang:2018yog,Martinez-Huerta:2020cut,Terzic:2021rlx,Addazi:2021xuf,Desai:2023rkd}. 

Despite the substantial body of work considering \ac{VC} in astrophysical environments, as part of electromagnetic cascades, there has been recently an increasing interest in using \ac{VC} radiation to put constraints on \ac{LIV} using air showers~\cite{rubtsov2017a, klinkhamer2017a, auger2022a, Duenkel:2023nlk}. If this occurs, even the detection of \ac{VHE} gamma rays may need to be re-evaluated, as they are observed through atmospheric effects, which could be influenced by \ac{LIV}.

The goal of this work is is to compute the spectra of \ac{VC} resulting from the propagation of \acl{HE} electrons and positrons in astrophysical environments~\cite{Kaufhold:2005vj, Anselmi:2011ae, Schreck:2017isa}:
\begin{equation}
e^{\pm} \rightarrow e^{\pm} + \gamma\,.
\end{equation}

\section{The kinematics of the vacuum Cherenkov effect}

We are considering the consequences of the resulting modified dispersion relations of the form
\begin{eqnarray}
&E_{e}^{2} = m_{e}^{2} + p_{e}^{2} + \sum\limits_{n=0}^{\infty} \chi_{n}^{e} \dfrac{p_{e}^{n+2}}{M_{\rm Pl}^{n}} \,, \label{EeDisp} \\[5pt]
&E_{\gamma}^{2} = k_{\gamma}^{2} + \sum\limits_{n=0}^{\infty}\chi_{n}^{\gamma} \dfrac{k_{\gamma}^{n+2}}{M_{\rm Pl}^{n}}\,, \label{EgammaDisp}
\end{eqnarray}
for electrons/positrons and photons, respectively.

In what follows, we will use the formalism and results from Ref.~\cite{Rubtsov:2012kb}\footnote{For pedagogical purposes and for future generalizations of this work, we change the notation with respect to Ref.~\cite{Rubtsov:2012kb} by expressing the \ac{LIV} parameters given there in terms of the ones defined by Eqs.~(\ref{EeDisp}) and (\ref{EgammaDisp}), i.e.~$\chi_{0}^{e} \equiv 2\kappa = 0$, $\chi_{2}^{e} \equiv 2g$, and $\chi_{2}^{\gamma} \equiv \xi$.}. We will restrict ourselves to the second-order case, with $n=2$, such that the only non-zero parameters are $\chi_{2}^{e}$ and $\chi_{2}^{\gamma}$, implying $f_{\rm LIV} \propto p^{4}$ in Eq.~(\ref{eq:generalDR}). This particular case is theoretically motivated, given that it corresponds to a \ac{mSME}~\cite{Colladay:1998fq}. 

\begin{figure}[ht]
    \centering
    \includegraphics[width=\columnwidth]{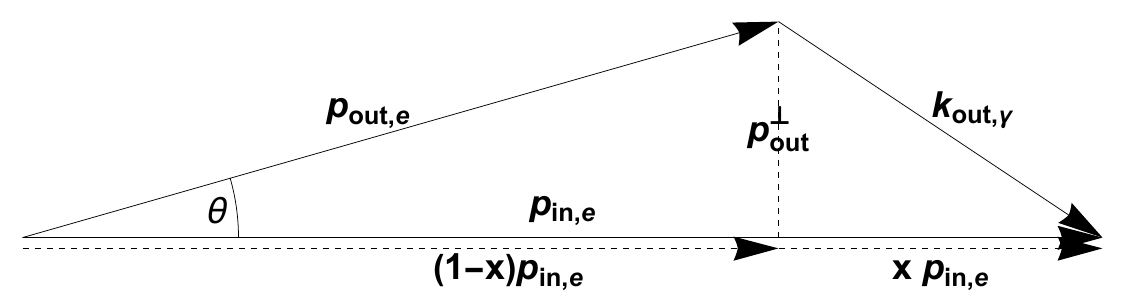}
    \caption{
    \ac{VC} kinetics formalism used in \cite{Rubtsov:2012kb} and in the present work.}
    \label{fig:VCkinematics}
\end{figure}

In the formalism of Ref.~\cite{Rubtsov:2012kb} (cf.~Fig.~\ref{fig:VCkinematics}), assuming the incoming electron with momentum $p_{{\rm in},e}$ to propagate in the positive $x_{1}$ direction, i.e.~$\mathbf{p}_{{\rm in},e} = p_{{\rm in},e} \mathbf{u}_{1}$, the outgoing electron has momentum $\mathbf{p}_{{\rm out},e} = p_{{\rm in},e}(1-x) \mathbf{u}_{1} + p_{\rm out}^{\perp} \mathbf{u}_{2}$, while the momentum vector $\mathbf{k}_{{\rm out},\gamma}$ of the outgoing photons is $\mathbf{k}_{{\rm out},\gamma} = x p_{{\rm in},e} \mathbf{u}_{1} - p_{\rm out}^{\perp} \mathbf{u}_{2}$. Here, $\mathbf{u}_{1}$ and $\mathbf{u}_{2}$ are two arbitrary unit vectors with $\mathbf{u}_{1} \perp \mathbf{u}_{2}$.

Conservation of energy and momentum implies~\cite{Rubtsov:2012kb}
\begin{equation} \label{omegapperp}
\frac{p_{\rm out,\perp}^{2}}{2 p_{{\rm in},e} x (1-x)} = \omega_{\rm LV}^{\rm VC}(x)\,,
\end{equation}
where
\begin{equation} \label{omegaLVVC}
\omega_{\rm LV}^{\rm VC}(x) = - \frac{\chi_{2}^{\gamma}}{2} \frac{p_{{\rm in},e}^{3} x^{3}}{M_{\rm Pl}^{2}} + \frac{\chi_{2}^{e}}{2} \frac{p_{{\rm in},e}^{3} \left(x^{3} - 3 x^{2} + 3 x\right)}{M_{\rm Pl}^{2}}\,,
\end{equation}
using which one can calculate two major quantities relevant for this work, namely, on the one hand, the \ac{VC} differential rate ${\rm d}\Gamma_{\rm VC}/{\rm d}x$ in terms of $x$ \cite{Rubtsov:2012kb},
\begin{equation} 
\dfrac{{\rm d}\Gamma_{\rm VC}}{{\rm d}x} = \alpha \left( \dfrac{2}{x} - 2 + x \right) \omega_{\rm LV}^{\rm VC}(x)\,,
\label{dGammadxVC}
\end{equation}
with $\alpha$ being the fine-structure constant, and, on the other hand, the deflection angle $\theta$, i.e.~the angle between the propagation directions of the incoming and the outgoing electrons, which from Fig.~\ref{fig:VCkinematics} and Eq.~(\ref{omegapperp}) can be seen to be
\begin{equation}
\theta = \arcsin\left[\sqrt{ \frac{2 x \omega_{\rm LV}^{\rm VC}(x)}{(1-x) p_{{\rm in},e} + 2 x \omega_{\rm LV}^{\rm VC}(x)} }\right]\,.
\label{eq:angleVCphotons}
\end{equation}

\section{Probabilities and Rates for the Vacuum Cherenkov Effect}

Ultimately, we wish to perform Monte Carlo simulations of electromagnetic cascades considering \ac{LIV}. For that, we require the spectrum of emission associated to the \ac{VC} effect. To this end, one has to consider the corresponding differential reaction rate ($\frac{{\rm d}\Gamma}{{\rm d}x}$), which is proportional to the differential probability ($\frac{{\rm d}P}{{\rm d}x}$) for the reaction to occur, ultimately resulting in a particle configuration with a given $x$ is proportional to it, given by the relation
\begin{equation} 
\frac{{\rm d}P}{{\rm d}x} = \frac{1}{\Gamma} \frac{{\rm d}\Gamma}{{\rm d}x}\,,
\label{diffGammaP}
\end{equation}
where $\Gamma$ is the total interaction rate. For a given configuration this expression is a function of a single free parameter for the configuration. Therefore, by determining $x$, the remainder of information follows immediately via the relations described below.

We start off by calculating the total photon emission rate for \ac{VC}, $\Gamma_{\rm VC}$, which is simply given by the integral of $\frac{{\rm d}\Gamma_{\rm VC}}{{\rm d}x}$ (see Eq.~(\ref{dGammadxVC})) over the allowed range of $x$,
\begin{equation}
\Gamma_{\rm VC} = \int\limits_\Upsilon \frac{{\rm d}\Gamma_{\rm VC}}{{\rm d}x} {\rm d} x \,,
\label{GammaVCintegral}
\end{equation}
with $\Upsilon = {\{x \; | \; 0 < x < 1 \; \wedge \; \omega_{\rm LV}^{\rm VC}(x) > 0\}}$.
The limits of integration, therefore, span the range for which $\omega_{\rm LV}^{\rm VC}$ is positive for $x \in (0; 1)$.

The allowed range for $x$ can be found analytically from the equation $\omega_{\rm LV}^{\rm VC} = 0$. When solving this equation, there are three different cases to be considered:

\noindent \textit{Case 1} ($\chi_{2}^{e} = \chi_{2}^{\gamma} = 0$). 
As here we have $\omega_{\rm LV}^{\rm VC}(x) = 0$ for the whole range of $x$, we simply get $\Upsilon = \{\}$, i.e.~no \ac{VC} is possible. This is plausible, as this case represents the situation in which Lorentz invariance is not broken and \ac{VC} does not occur.

\noindent \textit{Case 2} ($\chi_{2}^{e} = \chi_{2}^{\gamma} \neq 0$).
Here $\omega_{\rm LV}^{\rm VC}(x)$ reduces to a quadratic function of $x$ with $\omega_{\rm LV}^{\rm VC}(x) = 0$ being true for $x=0$ and $x=1$. Therefore, if $\chi_{e} > 0$, $\omega_{\rm LV}^{\rm VC}(x)$ can be represented by a concave-down parabola  and $\omega_{\rm LV}^{\rm VC}(x) > 0$ is true for the whole range of $x$, i.e.~$\Upsilon = (0;1)$. Meanwhile, for $\chi_{e} < 0$ we obtain a concave-up parabola, meaning $\omega_{\rm LV}^{\rm VC}(x) < 0$ for the relevant range and therefore rendering\ac{VC} impossible.

\noindent \textit{Case 3} ($\chi_{2}^{e} \neq \chi_{2}^{\gamma}$).
This is the most general case, where $\omega_{\rm LV}^{\rm VC}(x)$ is a cubic function of $x$. This generally yields three solutions for $\omega_{\rm LV}^{\rm VC}(x) = 0$, the trivial $x=0$, and two others, namely
\begin{equation} \label{omegaVCeq0}
x_{\rm VC,\pm} = - \frac{3 \chi_{2}^{e}}{2\left(\chi_{2}^{\gamma} - \chi_{2}^{e} \right)} \pm \frac{\sqrt{3 \chi_{2}^{\rm e} \left(4\chi_{2}^{\gamma} - \chi_{2}^{e} \right)}}{2 \left|\chi_{2}^{\gamma} - \chi_{2}^{e} \right|}\,,
\end{equation}
such that it requires a more detailed analysis which we briefly discuss in the following.

\begin{figure}[!ht]
    \centering      
    \includegraphics[width=1.1\columnwidth]{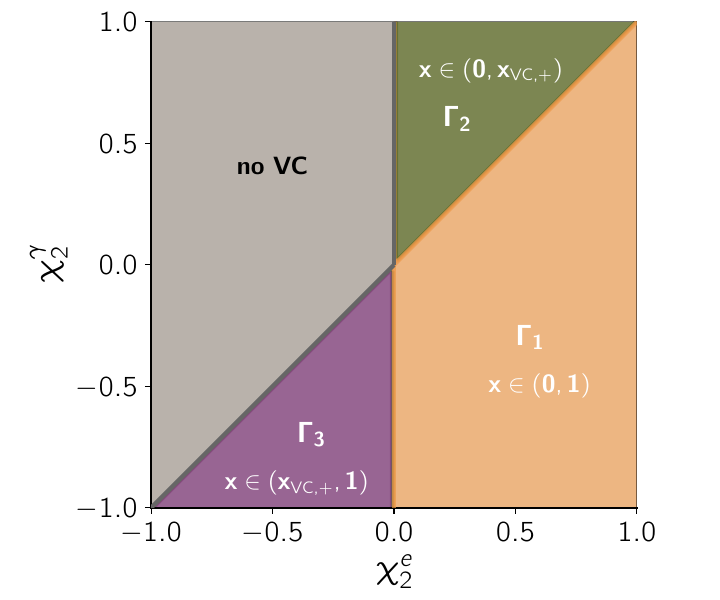}
    \caption{Overview of the intervals of $x$ for which the condition $\omega_{\rm LV}^{\rm VC}(x) > 0$ is fulfilled for a given combination of $\chi_{2}^{e}$ and $\chi_{2}^{\gamma}$. We identify several regions, according to the range of $x$ and the corresponding interaction rates ($\Gamma_i$).
    There are four distinct regions. The grey region, which also includes the case $(\chi_2^e = \chi_2^\gamma) \wedge  (\chi_2^\gamma < 0)$ and $(\chi_2^e = 0) \wedge (\chi_2^\gamma > 0)$, represents parameter combinations that result in $\omega_{\rm LV}^{\rm VC} \leq 0$ for $0 < x < 1$, meaning that \ac{VC} does not take place. The parameter combinations within the orange region correspond to $\omega_{\rm LV}^{\rm VC} > 0$ for all $x \in (0;1)$, with an emission rate $\Gamma_{\rm VC} = \Gamma_1$, following Eq.~(\ref{GammaVCcases}). Finally, for the purple and green regions, $\omega_{\rm LV}^{\rm VC} > 0$ for some part of the interval $(0;1)$, namely $(0;x_{\rm VC,+})$ and $(x_{\rm VC,+};1)$, respectively, where $x_{\rm VC,+}$ is given by Eq.~(\ref{omegaVCeq0}). The emission rate of \ac{VC} photons is also expected to be different in these regions. For combinations of $\chi_{2}^{e}$ and $\chi_{2}^{\gamma}$ corresponding to the gray region no \ac{VC} is possible. Note that the axes in this plot are \textit{linear}, while all the other region plots in this work are presented in \textit{symlog} axis scaling.}
  \label{fig:VCintlimitplot}
\end{figure}

First, one immediately sees that $x=0$ is the \textit{only} real solution if the radicand is \textit{negative}, i.e.~if $3 \chi_{2}^{\rm e} \left(4\chi_{2}^{\gamma} - \chi_{2}^{e} \right) < 0$, which is true for $\chi_{2}^{e} \gtrless 4 \chi_{2}^{\gamma}$ and $\chi_{2}^{e} \gtrless 0$, respectively. As a consequence, it follows that, for a negative radicand, we retrieve the full interval $(0;1)$ for $\chi_{2}^{\rm e}>0$, while for $\chi_{2}^{\rm e}<0$ \ac{VC} is not possible. If the radicand is \textit{non-negative}, we have to distinguish a number of different cases, an analysis the description of which goes beyond the scope of this work, but which in the end yields the combined domain of $x$:
\begin{equation} \label{xinVC}
x \in
\begin{cases}
(0;1)\,, & (\chi_{2}^{e} \ge 0) \wedge (\chi_{2}^{\gamma} \le \chi_{2}^{e})\,, \\
(0;x_{\rm VC,+}) \,, & 0 < \chi_{2}^{e} < \chi_{2}^{\gamma}\,, \\
(x_{\rm VC,+};1) \,, & \chi_{2}^{\gamma} < \chi_{2}^{e} < 0\,, \\
\{\} \,, & \text{else}\,.
\end{cases}
\end{equation}
The graphical representation of all these conditions is shown in Fig.~\ref{fig:VCintlimitplot}.

\begin{figure*}[!ht]
    \centering  
    \includegraphics[width=\columnwidth]{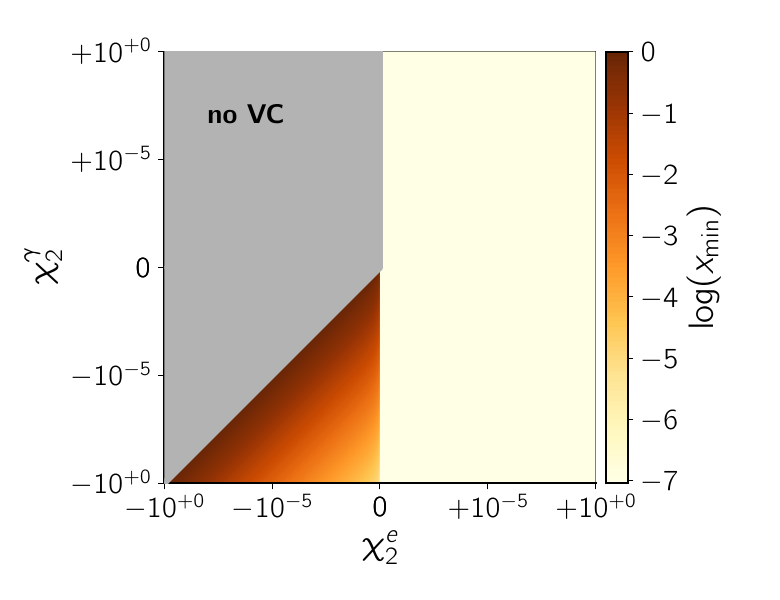} 
    \includegraphics[width=\columnwidth]{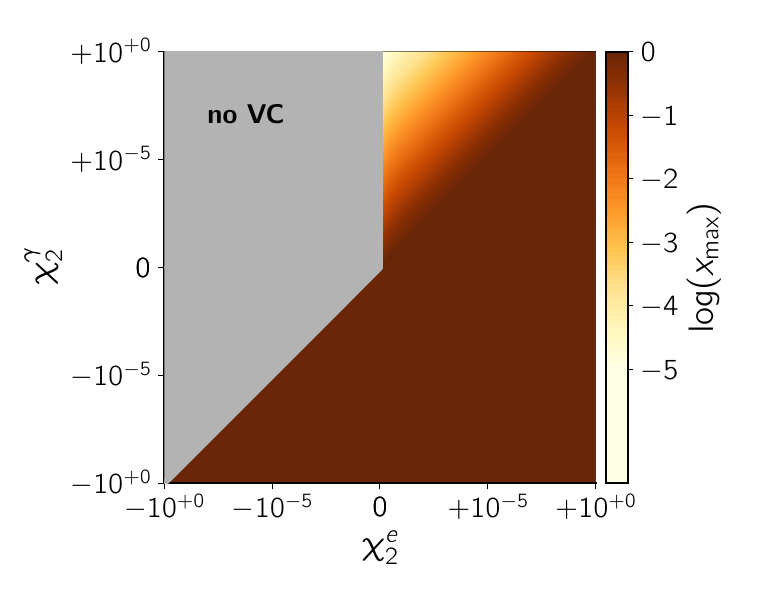}
    \caption{The lower (left panel) and upper (right panel) limits of integration according to Eq.~(\ref{xinVC}), i.e.~the limits for the possible values of $x$. The gray region corresponds to parameter combinations for which \ac{VC} is not possible. This and the following parameter-space plots are presented in \emph{symlog} scale, being linear between $-10^{-10}$ and $+10^{-10}$, and logarithmic outside this range.}
    \label{fig:VCintlimit2Dplot}
\end{figure*}

\begin{figure}[ht]
    \centering   
    \includegraphics[width=\columnwidth]{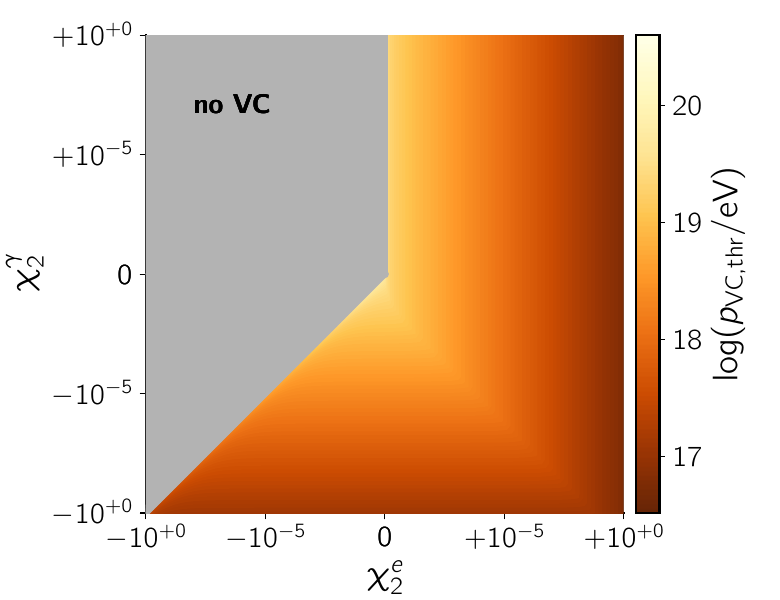}
    \caption{The threshold values $p_{\rm VC,thr}$ for \ac{VC} according to Eq.~(\ref{pVCthr}).}
    \label{fig:pVCthr}   
\end{figure}

A final point to be considered concerns the kinematical threshold for the reaction; \ac{VC} is only possible if the momentum of the incoming electron exceeds a threshold value $p_{\rm VC,thr}$, i.e.~$p_{\rm in,e} > p_{\rm VC,thr}$. For the case considered here, \ac{LIV} of order $n=2$, the only non-zero coefficients are $\chi_{2}^{e}$ and $\chi_{2}^{\gamma}$. In this case, as shown in~\cite{Jacobson:2002hd}, the threshold value can be written as
\begin{equation} \label{pVCthr}
\begin{split}
&\frac{p_{\rm VC,thr}}{\sqrt{m_{e} M_{\rm Pl}}} \\
&=\begin{cases}
\left( 3 \chi_{2}^{e} \right)^{-\frac{1}{4}} \,,\, & (\chi_{2}^{e} > 0) \wedge (\chi_{2}^{\gamma} \geq \beta \chi_{2}^{e})\,, \\
F(\lambda,\tau)^{-\frac{1}{4}} \,,\, & \left( \chi_{2}^{\gamma} < \beta \chi_{2}^{e} < 0 \right) \vee \left( \chi_{2}^{\gamma} < \chi_{2}^{e} \leq 0 \right) \,, \\
\infty \,, & (\chi_{2}^{e} < 0) \wedge (\chi_{2}^{\gamma} > \chi_{2}^{e})\,,
\end{cases}
\end{split}
\end{equation}
where
\begin{equation}
\begin{split}
&\beta = -(8 + 6\sqrt{2}) \,,\,
\lambda = \chi_{2}^{e} - \chi_{2}^{\gamma}\,,\, \tau = (\chi_{2}^{e} + 2\chi_{2}^{\gamma})/\lambda \,, \\ &F(\lambda,\tau) = \frac{2}{27} \lambda \left[ \tau^{3} + (\tau^{2} - 3)^{\frac{3}{2}} - \frac{9}{2}\tau \right]\,.
\end{split}
\end{equation}
The constraints on the threshold momentum in the $\chi_{2}^{e}$--$\chi_{2}^{\gamma}$ parameter space are shown in Fig.~\ref{fig:pVCthr}.

\begin{figure}[!ht]
    \centering  
    \includegraphics[width=\columnwidth]{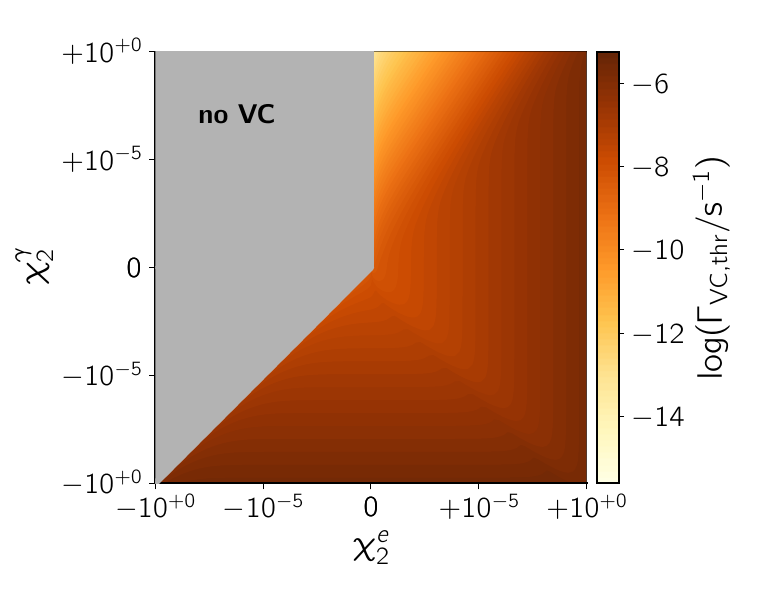}
    \caption{Magnitude of $\Gamma_{\rm VC,thr}$, i.e.~the total interaction rate for \ac{VC} at the \ac{VC} threshold for the considered parameter range. For combinations of $\chi_{2}^{e}$ and $\chi_{2}^{\gamma}$ corresponding to the gray region no \ac{VC} is possible.}
    \label{fig:GammaVCminPlot}
\end{figure}

The existence of a (lower) threshold together with the integration limits for $x$ may be used in specific cases to determine upper and lower momentum values $k_{\rm out,\gamma}$ and $p_{{\rm out},e}$ for the outgoing particles (s.~below).

With these results we now can calculate the total interaction rate for \ac{VC}, $\Gamma_{\rm VC}$, by integrating Eq.~(\ref{dGammadxVC}) in the range specified in Eq.~(\ref{xinVC}), for a given combination of $\chi_{2}^{e}$ and $\chi_{2}^{\gamma}$, obtaining
\begin{equation} 
\Gamma_{\rm VC} = 
\begin{cases}
\Gamma_1 \equiv \alpha \mathcal{G}_{0} \dfrac{p_{{\rm in},e}^{3}}{M_{\rm Pl}^{2}}  \,, & x \in (0;1)\,, \\
\Gamma_2 \equiv \alpha \mathcal{G}_{+} \dfrac{p_{{\rm in},e}^{3}}{M_{\rm Pl}^{2}}  \,, & x \in (0;x_{\rm VC,+})\,, \\
\Gamma_3 \equiv \alpha (\mathcal{G}_{0} - \mathcal{G}_{-})  \dfrac{p_{{\rm in},e}^{3}}{M_{\rm Pl}^{2}} \,, & x \in (x_{\rm VC,+};1)\,,
\end{cases}
\label{GammaVCcases}
\end{equation}
where
\begin{equation}
\mathcal{G}_{0} \equiv \dfrac{157 \chi_{2}^{e} - 22 \chi_{2}^{\gamma}}{120}  \,,
\end{equation}
and the parameters $\mathcal{G}_\pm$ are defined as
\begin{widetext}
\begin{equation} 
\mathcal{G_{\pm}} \equiv \frac{\chi_{2}^{e} \left( \pm S - 3 \chi_{2}^{e} \right) \left[ 37 \left( \pm \mathcal{S} - 6 \chi_{2}^{\gamma} \right) \chi_{2}^{e} \chi_{2}^{\gamma} - 64 \left(\chi_{2}^{e}\right)^{3} - \left( \pm 14 \mathcal{S} - 207 \chi_{2}^{\gamma} \right) \left(\chi_{2}^{e}\right)^{2} - 10 \left( \pm 5 \mathcal{S} - 16 \chi_{2}^{\gamma} \right) \left(\chi_{2}^{\gamma} \right)^{2} \right]}{160 \left( \chi_{2}^{\gamma} - \chi_{2}^{e} \right)^{4}}\,,
\end{equation}
\end{widetext}
wherein
\begin{equation}
\mathcal{S} \equiv \sqrt{3 \chi_{2}^{\rm e} \left(4\chi_{2}^{\gamma} - \chi_{2}^{e} \right)}\,.
\end{equation}
Employing these definitions, we can now understand the behavior of the \ac{VC} rate considering the $\chi_{2}^{e}$--$\chi_{2}^{\gamma}$ parameter space. The possible values of $\Gamma_{\rm VC}$ for all the different parameter combinations are shown in Fig.~\ref{fig:VCintlimitplot}.

Fig.~\ref{fig:GammaVCminPlot} indicates the numerical values of $\Gamma_{\rm VC,thr} = \Gamma_{\rm VC}(p_{{\rm in},e} = p_{\rm VC,thr})$ at the \ac{VC} threshold momentum. It is noticeable that the \ac{VC} rate is rather large, at least $\Gamma_{\rm VC,thr} \gtrsim 10^{11}\,{\rm Mpc^{-1}}$, which grows even further with $p_{{\rm in},e}$. In this energy range, the main Lorentz-invariant reactions competing with \ac{VC} for incoming electrons are \acl{ICS} and triplet pair production, whose interaction rates are below $100\,{\rm Mpc^{-1}}$~\cite{Batista:2021rgm}. Therefore, since competing processes are sub-dominant with respect to \ac{VC} by several orders of magnitude, we can affirm that, as long as $p_{{\rm in},e}$ lies above the threshold $p_{\rm VC,thr}$ (see above), \ac{VC} will occur on much shorter scales, typically below 1~\acl{A.U.}.

From Eq.~(\ref{GammaVCcases}), using Eq.~(\ref{diffGammaP}), we can finally calculate the differential probability with respect to the fraction $x$ of momentum carried away by the photon, which together with Eq.~(\ref{pVCthr}) gives
\begin{equation} \label{dPVCdxcases}
\dfrac{{\rm d}P_{\rm VC}}{{\rm d}x} = \max\left\{ 0, \,
  \alpha \left( \dfrac{2}{x} - 2 + x \right) \dfrac{\omega_{\rm LV}^{\rm VC}(x)}{\Gamma_{\rm VC}} \right\} \,,
\end{equation}
for $p_{{\rm in},e} > p_{\rm VC,thr}$ with $\Gamma_{\rm VC}$ given by Eq.~(\ref{GammaVCcases}). 

The first important conclusion from this is that the differential probability with respect to $x$ is independent of $p_{{\rm in},e}$ (as long as $p_{{\rm in},e} > p_{\rm VC,thr}$), such that a single distribution may be used for any electron participating in \ac{VC}. In order to identify parameter combinations which represent different regimes of \ac{VC}, we analyze the length of the intervals over which we carry out the integration, as this represents how narrow the distribution is located around its maximum, which may be seen in Fig.~\ref{fig:VCintlimit2Dplot}. 

Based on the symmetry of Fig.~\ref{fig:VCintlimit2Dplot}, we illustrate Eq.~(\ref{dPVCdxcases}) by exemplarily considering parameter combinations which obey the relation $\left| \chi_{2}^{\gamma}  \chi_{2}^{e} \right| = 10^{-10}$.

Finally, using the differential probability {\it with respect to $x$} presented in Eq.~(\ref{dPVCdxcases}), we can also calculate the differential probability {\it with respect to the deflection angle $\theta$},
\begin{equation} \label{dPdtheta}
\dfrac{{\rm d}P_{\rm VC}}{{\rm d}\theta} = \dfrac{{\rm d}P_{\rm VC}}{{\rm d}x} \dfrac{{\rm d}x}{{\rm d}\theta}\,,
\end{equation}
where ${\rm d}x/{\rm d}\theta$ can be calculated from Eq.~(\ref{eq:angleVCphotons}). This, in turn, enables us to calculate the average deflection angle $\bar{\theta}$ due to \ac{VC} to be
\begin{equation} \label{thetamean}
\bar{\theta} = \int \theta \dfrac{{\rm d}P_{\rm VC}}{{\rm d}x} \dfrac{{\rm d}x}{{\rm d}\theta} {\rm d}\theta\,.
\end{equation}

\section{Results}

In the following we present the \ac{VC} spectra for incoming electrons with initial energy $E_{{\rm in},e} = 10^{21}\,{\rm eV}$ and different \ac{LIV} parameter combinations. These have been obtained by employing a Monte Carlo approach with at least $10^{6}$ randomly selected samples obeying Eq.~(\ref{dPVCdxcases}) for each scenario.

As argued before, the rate of emission of photons ($\Gamma_{\rm VC}$), given by Eq.~(\ref{GammaVCcases}), is extremely short compared to the distance scales of interest (Galactic and cosmological distances). Therefore, we consider the emission of \ac{VC} photons to be an instantaneous process, such that we do not have to track the particles as they propagate. This approximation holds if the initial momentum of the electron ($p_{\rm in, e}$) is sufficiently large. Since we are interested in high-energy phenomena, electrons are relativistic and, consequently, the approximation is excellent.

As we will show in the following, we can identify four parameter combinations for which the emission spectra have specific qualitative behaviors. In order to simplify the further discussion, we label these ranges by as cases A to D and define them as
\begin{equation}
\begin{cases}
{\rm A}, & \chi_{2}^{\gamma} < \chi_{2}^{e} < 0\,, \\
{\rm B}, & \chi_{2}^{\gamma} < 0 \le \chi_{2}^{e}\,, \\
{\rm C}, & 0 < \chi_{2}^{\gamma} \le \chi_{2}^{e} \,, \\
{\rm D}, & 0 < \chi_{2}^{e} < \chi_{2}^{\gamma} \,, 
\end{cases}   
\label{eq:cases}
\end{equation}

\begin{figure*}[htb!]
    \includegraphics[width=\columnwidth]{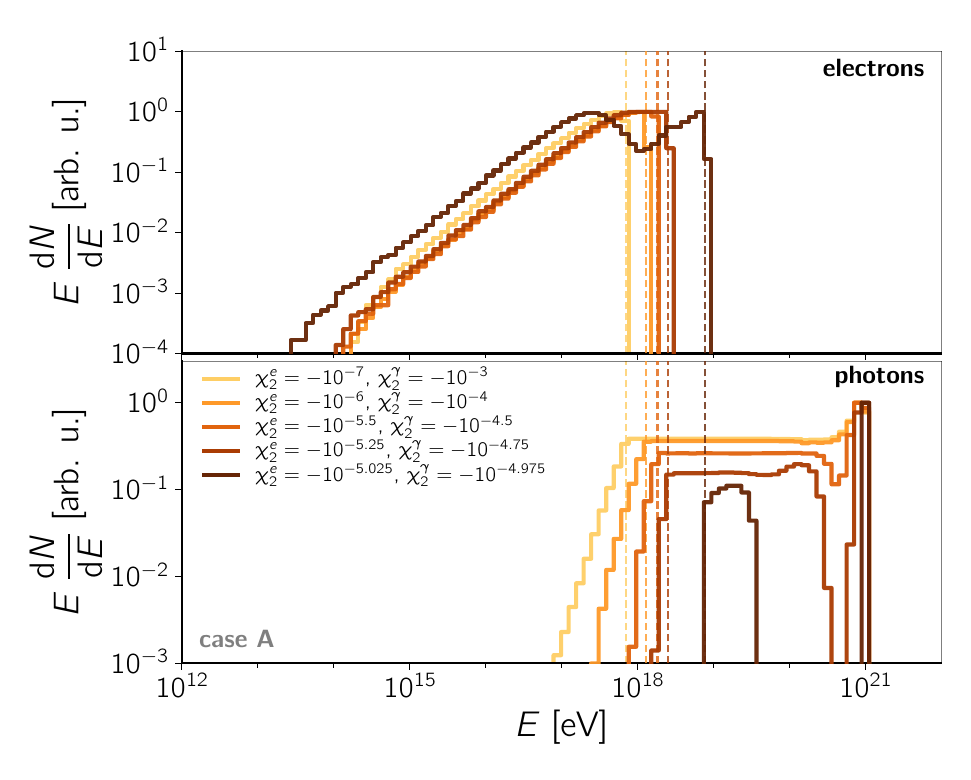}
    \includegraphics[width=\columnwidth]{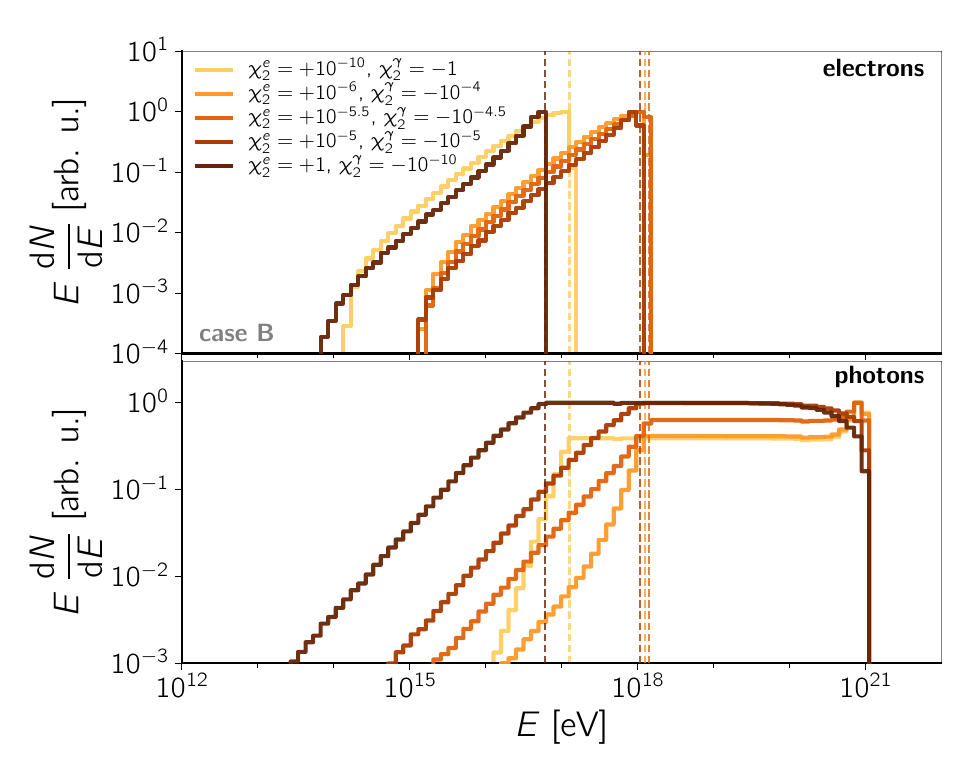}
    \includegraphics[width=\columnwidth]{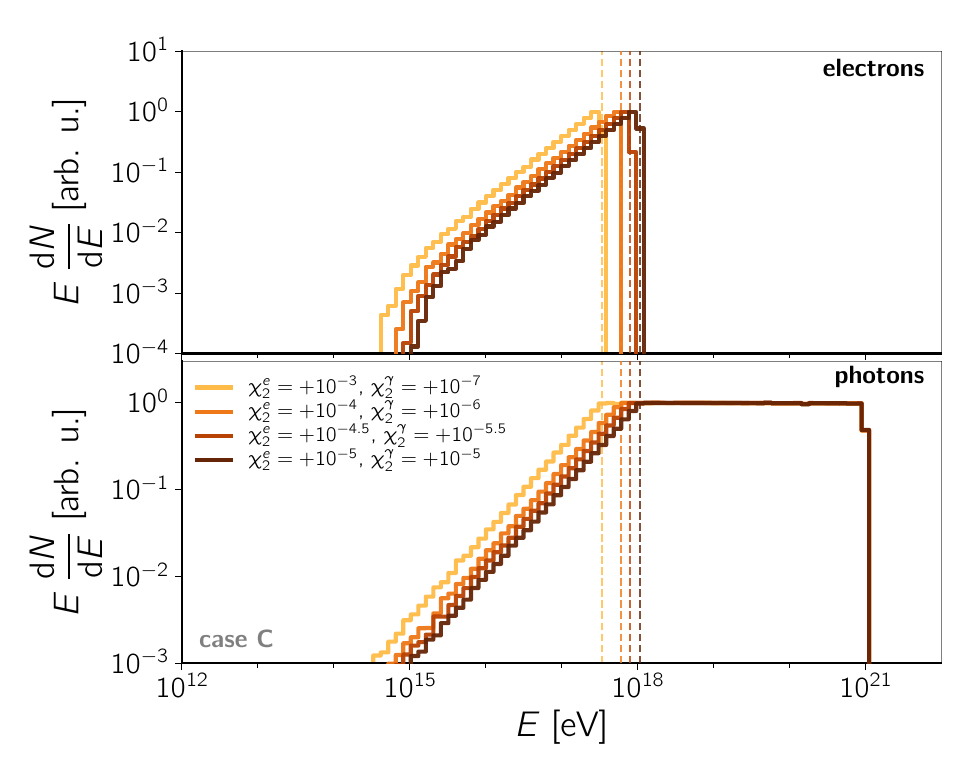}
    \includegraphics[width=\columnwidth]{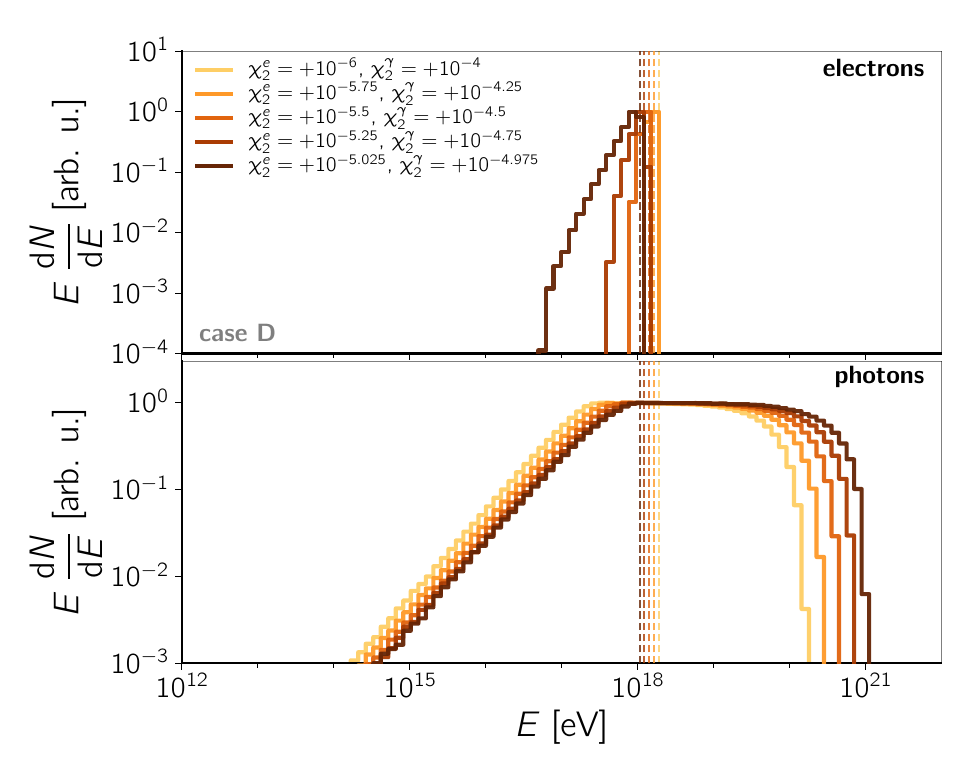}
    \caption{Numerical results obtained via Monte Carlo for various combinations of $\chi_2^e$ and $\chi_2^\gamma$. Dashed vertical lines indicate the threshold energies for the given combination of parameters. The vertical axis is
    weighted by  the corresponding energy to enhance the ${\rm d}N/{\rm d}E \propto E^{-1}$ behavior described in the text.
    The normalization of the distributions is arbitrary.}
    \label{fig:sim_a}
\end{figure*}

One important finding following directly from Fig.~\ref{fig:sim_a} is that, for certain parameter combinations, we observe a lower cut-off in the spectrum of one of the outgoing particles. This is evident for the electron spectra for \ac{LIV} parameters within range~D, and for the photon spectra for \ac{LIV} parameters within range~A.

\begin{figure*}[ht]
    \centering
    \includegraphics[width=\columnwidth]{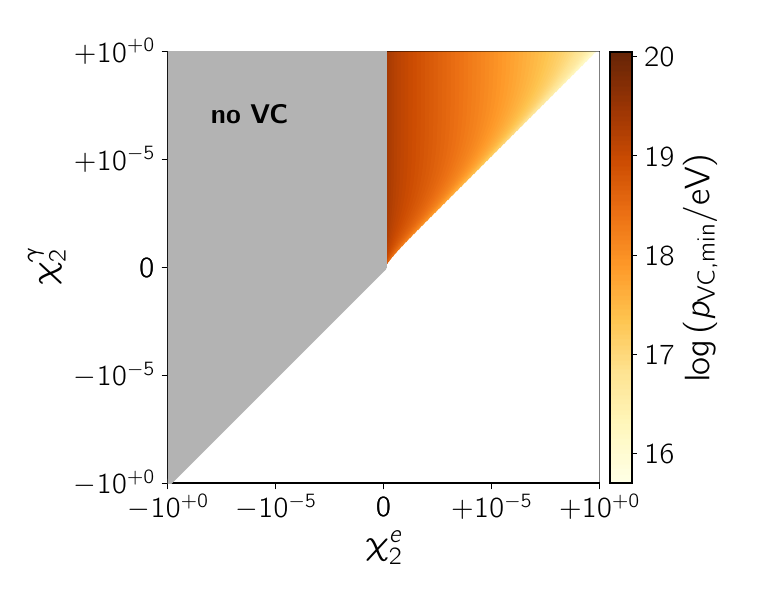}   \includegraphics[width=\columnwidth]{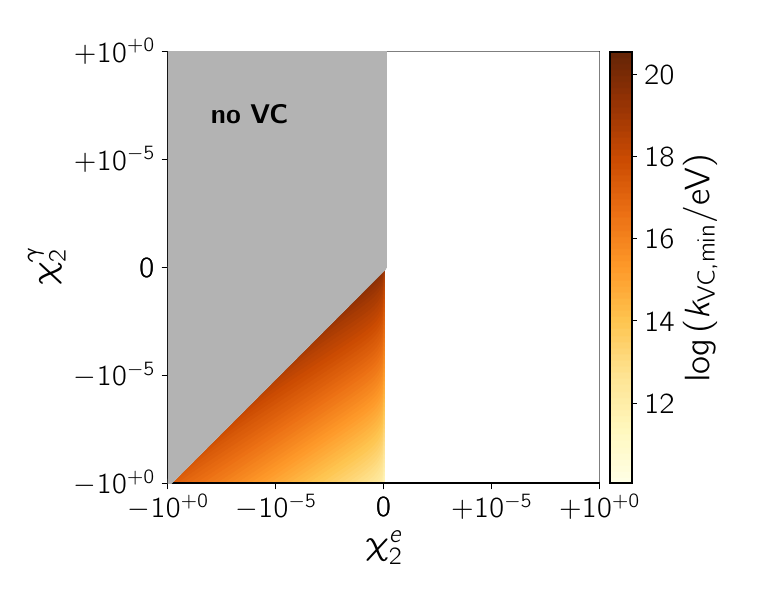}
    \caption{The lower limits for the momentum of $p_{\rm VC,min}$ for outgoing electrons (left) and $k_{\rm VC,min}$ for outgoing photons (right) from Eqs.~(\ref{pVCmin}) and (\ref{kVCmin}), respectively. The allowed regions may be found in these two equations as well. In the white regions a lower limit related to the threshold energy does not exist.}
    \label{fig:kpVCmin}   
\end{figure*}

This can be understood as follows. For the integration limits $x \in (0; x_{\rm VC,+})$, which corresponds to the condition $0 < \chi_{2}^{e} < \chi_{2}^{\gamma}$ according to Eq.~(\ref{xinVC}), we have $x < x_{\rm VC,+}$. As a consequence, for $(1 - x)$, which corresponds to the fraction of the incoming momentum carried away by the outgoing electron, we find that $1 - x > 1 - x_{\rm VC,+}$. Given that the lowest possible value for $p_{{\rm in},e}$ is determined by $p_{\rm VC,thr}$, we can write the relation as
\begin{equation} \label{pVCmin}
\begin{split}
p_{\rm VC,min} \le p_{{\rm out},e} \le p_{\rm VC,thr}
\end{split}
\end{equation}
for $0 < \chi_{2}^{e} < \chi_{2}^{\gamma}$, where $p_{\rm VC,min}$ is defined as
\begin{equation}
p_{\rm VC,min} \equiv (1-x_{\rm VC,+}) p_{\rm VC,thr}\,.
\end{equation}
On the other hand, for the integration limits $x \in (x_{\rm VC,+};1)$, i.e.~for $\chi_{2}^{e} < \chi_{2}^{\gamma} < 0$ according to Eq.~(\ref{xinVC}), we have $x \ge x_{\rm VC,+}$ and hence $1-x < 1-x_{\rm VC,+}$. Using the same reasoning as above, we can write down the relation
\begin{equation} \label{kVCmin}
k_{{\rm out},\gamma} \ge k_{\rm VC,min}
\end{equation}
for $\chi_{2}^{\gamma} < \chi_{2}^{e} < 0$, where $k_{\rm VC,min}$ is defined as
\begin{equation}
k_{\rm VC,min} \equiv x_{\rm VC,+} p_{\rm VC,thr}\,.
\end{equation}
The results of Eqs.~(\ref{pVCmin}) and (\ref{kVCmin}) are presented in Fig.~\ref{fig:kpVCmin}.

\begin{figure*}[htb!]
    \includegraphics[width=\columnwidth]{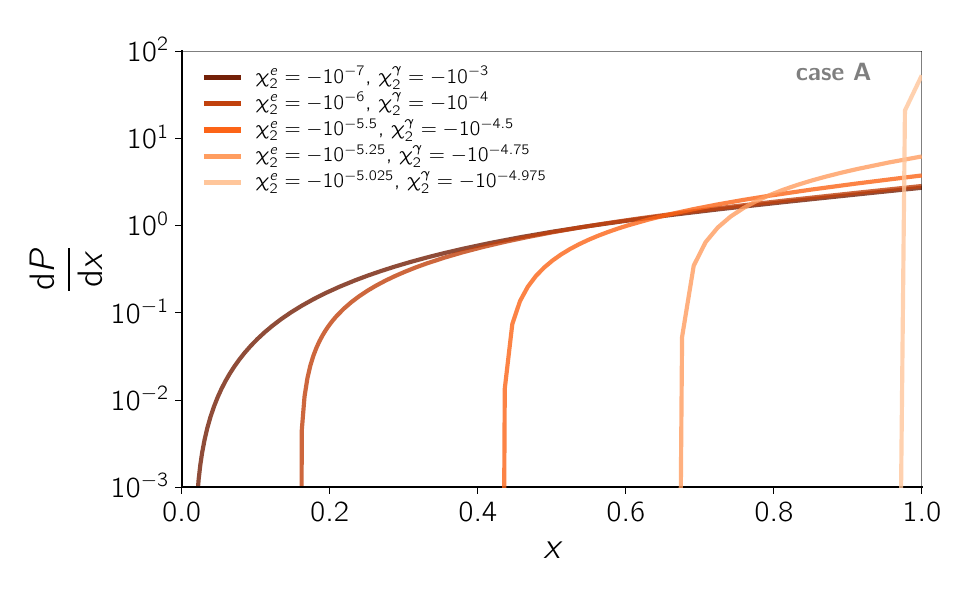}
    \includegraphics[width=\columnwidth]{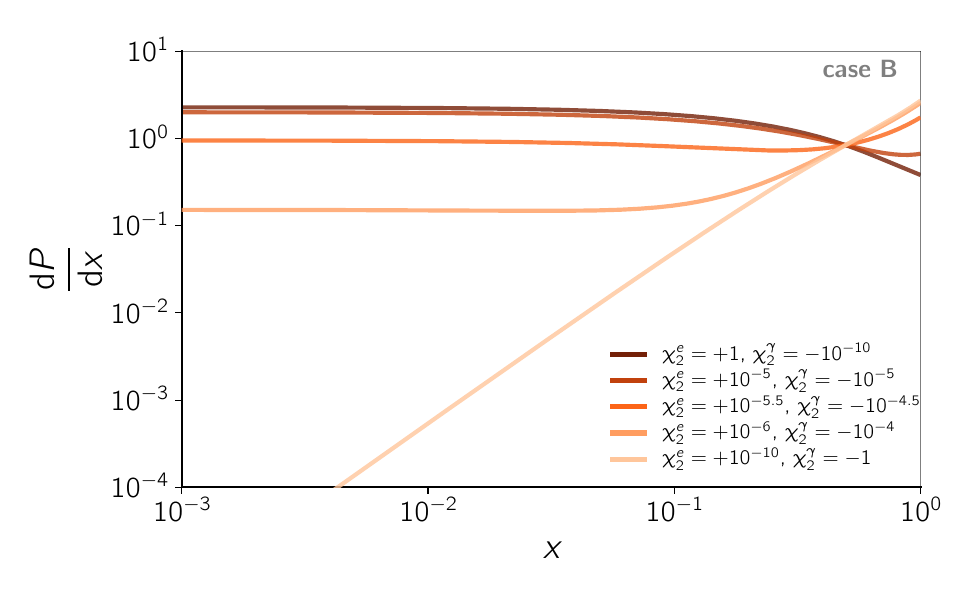}
    \includegraphics[width=\columnwidth]{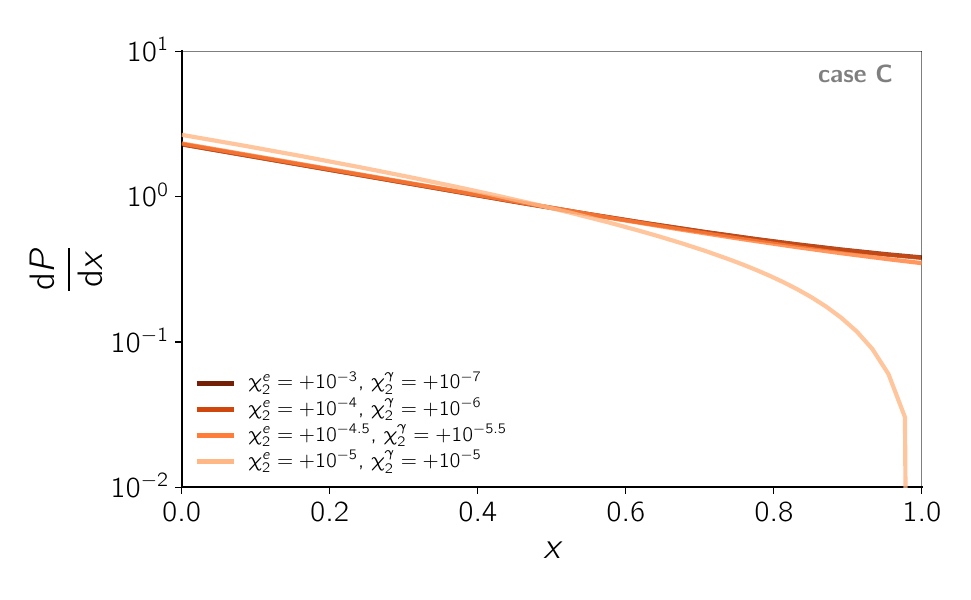}
    \includegraphics[width=\columnwidth]{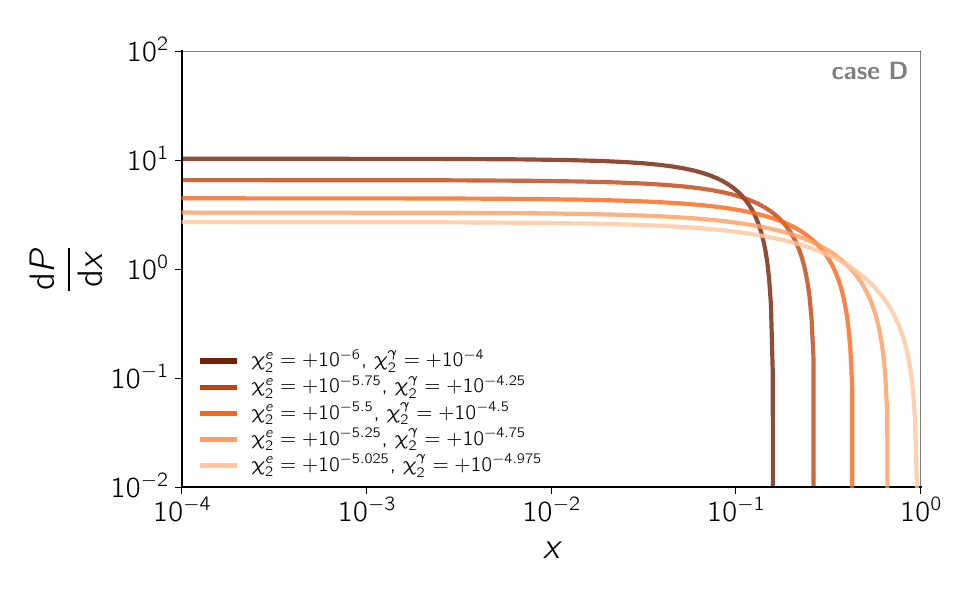}
    \caption{The plots show the differential probability distribution ($\frac{{\rm d}P_{\rm VC}}{{\rm d}x}$) for different combinations of $\chi_{2}^{e}$ and $\chi_{2}^{\gamma}$ satisfying the constraint $\left| \chi_{2}^{\gamma}  \chi_{2}^{e} \right| = 10^{-10} $. Each panel corresponds to different regimes, based on the signs of $\chi_{2}^{\gamma}$ and $\chi_{2}^{e}$, according to Eq.~(\ref{eq:cases})}.
    \label{fig:dPdx}
\end{figure*}

Another prominent feature, common to all photon spectra, is that all of them follow a particular power law\footnote{While the \textit{momentum} ($k$ and $p$) is the relevant \textit{kinematic} quantity, we discuss the spectra of Fig.~\ref{fig:sim_a} in terms of the \textit{energy}, which is the relevant \textit{observable} quantity. This is justified because for very high energies the approximation $E \simeq p$ still holds, even in the presence of \ac{LIV}, following the modified dispersion relations given by Eqs.~(\ref{EeDisp}) and (\ref{EgammaDisp}).}, namely ${\rm d}N/{\rm d}E_{{\rm out},\gamma} \propto E_{{\rm out},\gamma}^{-1}$, roughly from the \ac{VC} threshold value up to approximately the maximal energy of the initial electron, as shown in the corresponding panels of Fig.~\ref{fig:sim_a}. This is because the differential interaction probability does not depend on the momentum of the incoming electron, according to Eq.~(\ref{dPVCdxcases}).

To better understand this, we now consider an idealized scenario wherein an incoming electron with momentum $p_{{\rm in},e}$ emits a photon with momentum $k_{{\rm out},\gamma} = x_{0} p_{{\rm in},e}$ (with a fixed value of $x_{0}$ within the range $0<x_{0}<1$), such that the outgoing electron has momentum $p_{{\rm out},e} = (1-x_{0}) p_{{\rm in},e}$. This implies that the momentum of the $j$th photon emitted by the electron is
\begin{equation}
k_{{\rm out},\gamma}^{(j)} = x_{0} (1-x_{0})^{j-1} p_{{\rm in},e}\,.
\end{equation}
This happens as long as the momentum of the electron exceeds the threshold momentum, $p_{\rm VC,thr}$, or conversely, $(1-x_{0})^{n_{\gamma}} p_{{\rm in},e} > p_{\rm VC,thr}$, from which one can obtain the total number of emitted photons $n_{\gamma}$. This set of $n_{\gamma}$ photons corresponds to a distribution proportional to $k_{{\rm out},\gamma}^{-1}$, thus explaining the spectral features described above, as it applies even in the more complicated case of the actual \ac{VC} spectrum, shown in Fig.~\ref{fig:sim_a}.

We now turn to the different cases of the photon spectra, starting with a feature common to the cases C and D. While for higher energies one can see the power-law behavior ${\rm d}N/{\rm d}E_{{\rm out},\gamma}$ of proportional to $E_{{\rm out},\gamma}^{-1}$ as described above, the spectrum flattens to ${\rm d}N/{\rm d}E_{{\rm out},\gamma} \propto E_{{\rm out},\gamma}^{0}$ for lower energies. This is an immediate consequence of the differential probability distribution (see Eq.~(\ref{dPVCdxcases})) being peaked around zero and then falling roughly linearly for higher values of $x$, as shown in the corresponding panels of Fig.~\ref{fig:dPdx}. The transition between these two regimes is determined by the \ac{VC} threshold value.

Next, we examine the photon spectra for case A. In this case, as mentioned earlier, the spectrum exhibits a lower cut-off. Additionally, as the values of $\chi_{2}^{\gamma}$ and $\chi_{2}^{e}$ get closer to each other, a gap appears in the spectrum near the initial momentum, which widens as the two values converge, as supported also by the behavior of the case with $\chi_{2}^{e} = -10^{-5.025}$ and $\chi_{2}^{\gamma} = -10^{-4.975}$.

This gap-like feature is once more a direct consequence of the corresponding differential probability distributions (cf.~Fig.~\ref{fig:dPdx}). Here we can see that the closer the two values are to each other the narrower the differential distribution, which peaks around $1$, becomes. This produces a narrow peak in the photon spectrum close to the initial momentum value and then another peak before the electron drops below the threshold value. Once the values get further apart, the distribution widens, which itself results in a widening of the peak as well as its shift to lower momentum values.

Finally, we consider case B. Here one can see two different regimes for ${\rm d}N/{\rm d}E_{{\rm out},\gamma}$ at lower momentum values of the outgoing photons -- either a quadratic increase for $\left| \chi_{2}^{\gamma} \right| \gg \chi_{2}^{e}$ or a flat spectrum for $\left| \chi_{2}^{\gamma} \right| \ll \chi_{2}^{e}$. Note that for the quadratic case, for very small momenta, a shift to a flat spectrum might still take place, as can be seen for the parameter combination $\chi_{2}^{e} = 10^{-4}$ and $\chi_{2}^{\gamma} = -10^{-6}$.

This behavior may again be explained by analyzing the corresponding plots in Fig.~\ref{fig:dPdx}. Both regimes correspond directly to the behavior of the corresponding differential probability densities, as due to the fact that they are independent of the momentum of the incoming electron, the spectrum is effectively a superposition of many of such individual distributions. And indeed, looking at the extreme case with $\chi_{2}^{e}=10^{-10}$ and $\chi_{2}^{\gamma}=-1$, the quadratic term is dominating within the region given by $10^{-5} \lesssim x \lesssim 10^{-1}$. The transition between the two regimes takes place roughly when the differential probability distribution is symmetric.

Proceeding to the electron spectra, we see that they have significantly fewer notable features compared to the corresponding photon spectra. In particular, considering all the parameter combinations presented in Fig.~\ref{fig:sim_a} for case A, we see that the spectral shape of ${\rm d}N/{\rm d}E_{{\rm out},e}$ is a flat spectrum with a sharp cut-off which occurs when the electron momentum drops below the \ac{VC} threshold value. The only spectral variation appears right below the cut-off where the spectrum either slightly rises or falls. This is, again, a direct consequence of the corresponding differential probability distribution function, but this time resulting in a reverse relationship compared to the photon spectra: for a rising distribution the spectrum falls close to the cut-off (and vice versa). This can be easily understood by noting that the electron carries away the remaining fraction of the incoming momentum, $1-x$.

In contrast, we see that for the case with $\chi_{2}^{e} = -10^{-5.025}$ and $\chi_{2}^{\gamma} = -10^{-4.975}$ the spectrum has a step-like shape. As shown in Fig.~\ref{fig:VCintlimitplot},  the probability for an electron to obtain a given fraction of the incoming momentum is highly peaked around $0$, which means that the momentum of the outgoing electron is several orders of magnitude below the momentum of the incoming one. This difference is responsible for the aforementioned step-like feature as in this case, the electron emits one photon and loses energy, but remains above the threshold. It is only after a second photon is emitted that the electron's energy drops below the threshold, thus resulting in the aforementioned extremely narrow-peaked photon spectrum.

The other case for which the electron spectrum displays some interesting features (apart from the lower cut-off described above) is for \ac{LIV} parameter values lying within the ranges denoted as case D. If $\chi_{2}^{e}$ and $\chi_{2}^{\gamma}$ are fairly apart from each other, one gets an almost monochromatic emission due to the differential probability shown in the top panel of Fig.~\ref{fig:dPdx}. We can see that for electrons it is tightly peaked around $1$, which translates into the monochromatic spectrum. By further emitting photons, this distribution simply moves to lower momentum values until it reaches the threshold value at which point it does not change anymore. Once the values of $\chi_{2}^{e}$ and $\chi_{2}^{\gamma}$ are getting close to each other, this nearly monochromatic spectrum  acquires a low-momentum tail, again simply resulting directly from the now wider differential probability distribution.

\begin{figure}[htb!]
    \centering    \includegraphics[width=\columnwidth]{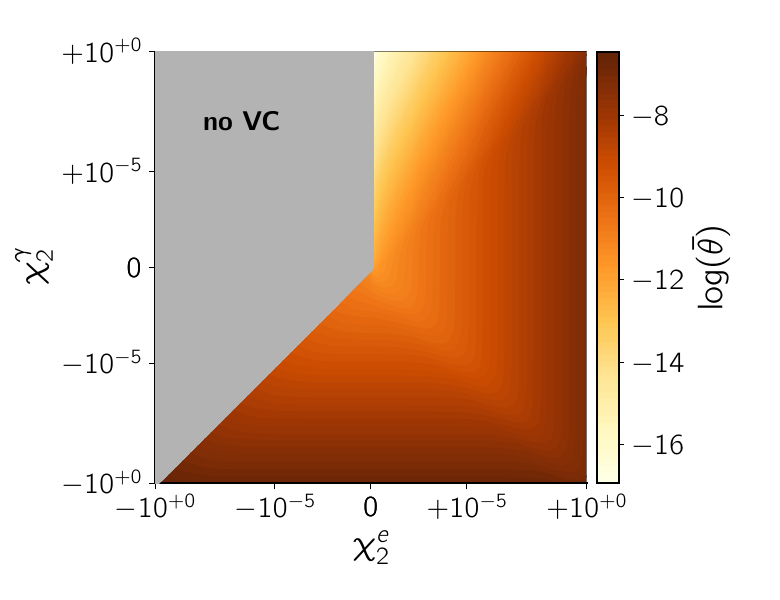}
    \caption{The average deflection angle $\bar{\theta}$ (in radians) according to Eq.~(\ref{thetamean}), for $p_{{\rm in},e} = 10^{21}\,{\rm eV}$.}
    \label{fig:angleplot}
\end{figure}

\bigskip
As a final point, we present our results regarding the angular distribution defined in Eq.~(\ref{dPdtheta}) or, more specifically, its main observable, the average deflection angle $\bar{\theta}$ from Eq.~(\ref{thetamean}). For the parameter space used in this work the latter is shown Fig.~\ref{fig:angleplot}.

As one can see, the overall behavior may be described as being opposite for $\chi_{2}^{e}$ and $\chi_{2}^{\gamma}$, meaning that while for $\chi_{2}^{e}$ the value of $\bar{\theta}$ is large for larger values of $\chi_{2}^{e}$, getting smaller the smaller $\chi_{2}^{e}$ gets, the relation is reversed for $\chi_{2}^{\gamma}$. Another important observation is that $\bar{\theta}$ increases linearly with $p_{{\rm in},y}$.

Both of these behaviors can be qualitatively explained by expanding the formula for $\theta$, Eq.~(\ref{eq:angleVCphotons}), in terms of $p_{{\rm in},e}/M_{\rm Pl}$, giving
\begin{equation}
\begin{split}
\theta &= \sqrt{\frac{x \left[ - \chi_{2}^{\gamma} x^{3} + \chi_{2}^{e}  \left(x^{3} - 3 x^{2} + 3 x\right)\right]}{1-x}} \left(\frac{p_{{\rm in},e}}{M_{\rm Pl}}\right) \\
&+ \mathcal{O}\left[ \left(\frac{p_{{\rm in},e}}{M_{\rm Pl}}\right)^{3}\right]
\end{split}
\end{equation}
Here, one can see the two behaviors described above. On the one hand, the opposite signs in front of $\chi_{2}^{e}$ and $\chi_{2}^{\gamma}$ and, on the other hand, the clear linear dependence on $p_{{\rm in},e}$.

\section{Discussion}

A particularly intriguing prospect is the potential to detect the \ac{VC} effect through astrophysical data. Our results point to some interesting signatures that could be observed using, for instance, gamma-ray observations.

One process that arguably competes with \ac{VC} is synchrotron emission. In the Milky Way, whose typical magnetic field is $B \sim 0.1 \; \text{nT}$, the energy-loss time of \ac{HE} electrons via \ac{VC} emission is over a dozen of orders of magnitude shorter than that of synchrotron, implying that the former completely dominates, for most of the parameter space investigated in this work. Note, however, that in highly magnetized environments, such as the surrounding of compact objects, this is no longer true, especially considering the $B^2$ dependence of the synchrotron emission time. However, one should also keep in mind that the synchrotron power in Lorentz-violating electrodynamics may change due to the modified dispersion relations, making this comparison non-trivial~\cite{altschul2005a, montemayor2005a, altschul2006a, zhukvosky2008a}.

\Acl{ICS} of \ac{HE} electrons interacting with background photons of the \ac{CMB}, \ac{EBL}, and \ac{CRB}, for instance, also compete with \ac{VC} emission. But in this case, \ac{ICS} cooling is much more inefficient than \ac{VC} for most of the parameter space explored.

We have analyzed the \ac{VC} effect only for electrons and positrons, but all results can be generalized in a straightforward manner to other charged leptons, namely muons and tauons provided, of course, that the lifetimes of these particles in the Earth frame is not much shorter than the corresponding \ac{VC} time scale. This is because our treatment was performed within the framework of \ac{QED}.

The formalism can probably be extended to other charged \emph{elementary} particles. 
However, it is far from clear if the same applies to composite particles such as protons. For further details on this debate, the reader is referred to works discussing the so-called `soccer ball problem'~\cite{amelinocamelia2011a, hossenfelder2013b, amelinocamelia2013d, hossenfelder2014a, amelinocamelia2017c, kumar2020a}.

Considering the energy-dependent speed of light arising in \ac{LIV} models, there could be an interesting interplay between the \ac{VC} photon spectrum and its temporal dependence. This would be particularly useful for \ac{QG} searches using electromagnetic observations of, for example, gamma rays, such as those reviewed in Ref.~\cite{Addazi:2021xuf}.
This promising avenue~\cite{alvesbatista2025a} would require a comprehensive treatment of all relevant astrophysical processes, such as the one started in Ref.~\cite{Saveliev:2023urg}.

A central aspect of our work is the independent treatment of the \ac{LIV} coefficients for different species of particles. This approach is particularly important for future analyses that aim to bring together distinct facets of \ac{LIV}-related phenomenology across various particles. Specifically, due to \ac{CPT} symmetry, the LIV coefficients associated with \ac{CPT}-odd operators do have opposite signs for particles and their corresponding antiparticles~\cite{altschul2007c, antonelli2020a, Addazi:2021xuf}. This sign difference leads to distinct observational signatures for particles and antiparticles, which is crucial for experimental searches and interpretations.

\section{Conclusions and Outlook}

In this work, we presented the first comprehensive derivation of the \acl{VC} radiation spectrum, based on the interaction rates from \cite{Rubtsov:2012kb}.

The most significant phenomenological result is that the simplified ``binary'' approach to \ac{VC}, used in works like Ref.~\cite{Saveliev:2023urg}), where the electron momentum is reduced to the \ac{VC} threshold and the emitted photon momentum is simply the difference between the incoming and outgoing electron, proves to be an oversimplification in most cases. Although there are instances where the outgoing electrons are nearly monochromatic (see Fig.~\ref{fig:sim_a}), this is not generally true for the emitted photons. Thus, we conclude that one must generally consider the full \ac{VC} spectrum rather than relying on simplified assumptions.

Given the short interaction time derived in Eq.~(\ref{fig:GammaVCminPlot}), \ac{VC} radiation occurs almost immediately after the electron is created compared to the distance scales of interest (Galactic distances and beyond). Therefore, these spectra can serve as effective inputs for subsequent processes, triggering the development of electromagnetic cascades. Indeed, in the future we will implement this treatment and model them employing the CRPropa simulation framework~\cite{alvesbatista2016a, AlvesBatista:2022vem}. 

Another aspect to explore in future works is the effect of \ac{LIV} on the spectra and angular distributions resulting from other processes relevant to electromagnetic cascades, such as \acl{ICS}, \acl{PP}, and \acl{PD}, as they may be used to distinguish reactions and to establish limits on the \ac{LIV} parameters. In particular, there are parameter combinations of $\chi_{2}^{e}$ and $\chi_{2}^{\gamma}$ where both \acl{VC} and \acl{PD} can occur, thus warranting further detailed investigation.

\begin{acknowledgments}
The work of AS is supported by the Russian Science Foundation under grant no.~22-11-00063.
RAB acknowledges the support of the Agence Nationale de la Recherche (ANR), project ANR-23-CPJ1-0103-01. 
\end{acknowledgments}

\bibliography{bib.bib}

\end{document}